\documentclass[aps,prb,twocolumn,superscriptaddress]{revtex4-2}

\usepackage{graphicx,bm,color}
\usepackage[bbgreekl]{mathbbol}

\newcommand{\ud}{{\uparrow\downarrow}}
\newcommand{\du}{{\downarrow\uparrow}}
\newcommand{\uu}{{\uparrow\uparrow}}
\newcommand{\dd}{{\downarrow\downarrow}}
\newcommand{\si}{\sigma}
\newcommand{\sib}{{\bar{\sigma}}}

\newcommand{\bfm}{{\bf m}}

\newcommand{\SC}{{\cal S}}

\newcommand{\bigO}{\mathcal{O}}

\newcommand{\kf}{k_F}
\newcommand{\ef}{\varepsilon_F}
\newcommand{\bfk}{{\bf k}}
\newcommand{\bfkp}{{\bf k'}}
\newcommand{\kp}{k'}
\newcommand{\kpsq}{{\kp}^2}
\newcommand{\zt}{\tilde z}
\newcommand{\zst}{z_{\s\tp}}

\newcommand{\bfq}{{\bf q}}

\newcommand{\bfr}{{\bf r}}
\newcommand{\bfrho}{{\bm{\rho}}}
\newcommand{\bfrp}{{\bfr{'}}}
\newcommand{\bfs}{{\bf s}}
\newcommand{\bp}{b'}
\newcommand{\tp}{\tau}
\newcommand{\s}{\sigma}
\newcommand{\w}{\omega}

\newcommand{\ua}{\uparrow}
\newcommand{\da}{\downarrow}

\newcommand{\ku}{k_{F\ua}}
\newcommand{\kd}{k_{F\da}}
\newcommand{\ks}{k_{F\si}}
\newcommand{\ksb}{k_{F\sib}}

\newcommand{\essb}{\varepsilon_{\si\sib}}

\newcommand{\z}{\zeta}

\newcommand{\en}[2]{\varepsilon_{#1}(#2)}

\begin{document}

\title{Spin waves in doped graphene: a time-dependent spin-density-functional approach to
collective excitations in paramagnetic two-dimensional Dirac fermion gases}

\author{Matthew J. Anderson}
\affiliation{Department of Physics and Astronomy, University of Missouri, Columbia, Missouri 65211, USA}

\author{Florent Perez}
\affiliation{Institut des Nanosciences de Paris, CNRS/Universit\'e Paris VI, Paris 75005, France}

\author{Carsten A. Ullrich}
\affiliation{Department of Physics and Astronomy, University of Missouri, Columbia, Missouri 65211, USA}

\date{\today }

\begin{abstract}
In spin-polarized itinerant electron systems, collective spin-wave modes arise from dynamical exchange and correlation (xc) effects.
We here consider spin waves in doped paramagnetic graphene with adjustable Zeeman-type band splitting. The spin waves
are described using time-dependent spin-density-functional response theory, treating dynamical xc effects within the
Slater and Singwi-Tosi-Land-Sj{\"o}lander approximations. We obtain spin-wave dispersions and spin stiffnesses as a function
of doping and spin polarization, and discuss prospects for their experimental observation.
\end{abstract}

\maketitle

\section{Introduction}\label{sec1}

Graphene is a material with many fascinating structural and electronic properties \cite{Neto2009}.
Most notably, it has a Dirac cone feature in its energy bands that leads to the electrons behaving as massless particles.
In its pristine form, graphene is a semimetal; it can be made metallic through doping or gating, which then leads to a wealth
of applications, for instance in plasmonics \cite{Koppens2011,Grigorenko2012,Luo2013,Huang2017}. Together with many other new 2D materials, graphene also shows promise
for spintronics applications \cite{Avsar2020}.

\begin{figure}
  \includegraphics[width=\linewidth]{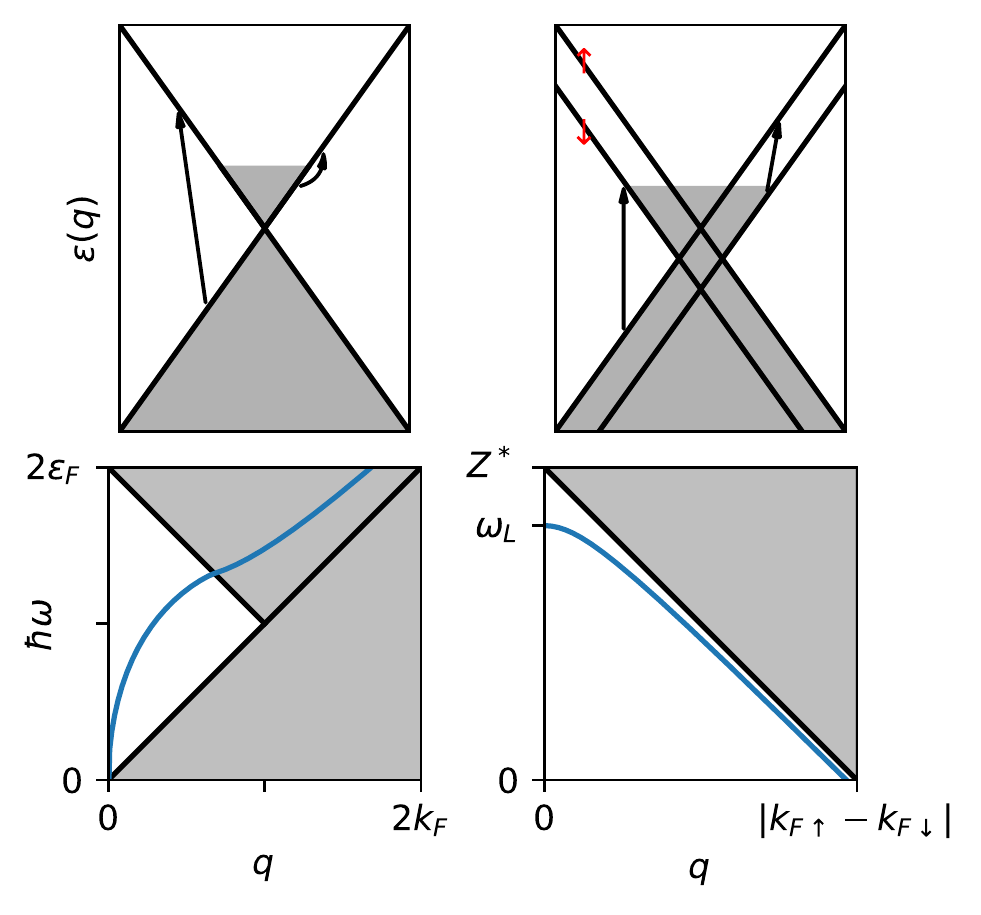}
  \caption{Top left: spin-conserving single-particle transitions of doped, nonmagnetic graphene near a Dirac point.
  Bottom left: associated plasmon dispersion and intra- and interband single-particle continua.
  Top right: spin-flip transitions of doped, magnetized graphene near a Dirac point. The bands are split by the Zeeman energy $Z^*$.
  Bottom right: associated single-particle spin-flip continuum and spin-wave dispersion, where $\omega_L$ denotes the Larmor frequency. }
  \label{fig1}
\end{figure}

Plasmons are collective charge-density excitations, which can be characterized as the collective response of an electron gas
to an induced electrostatic perturbation. Typically, plasmon mode frequencies and dispersions are calculated using the random-phase approximation (RPA);
such calculations were done for graphene early on \cite{Wunsch2006,Hwang2007,Abedinpour2011,Stauber2014,Agarwal2015}.
The left panels of Fig. \ref{fig1} give a schematic illustration of plasmons in doped graphene, showing inter- and intraband single-particle excitations
and the plasmon dispersion. The latter is very similar to the plasmon dispersion in a two-dimensional electron gas (2DEG) \cite{GiulianiVignale}, with a  characteristic
$\sqrt{q}$ behavior for small wavevectors $q$. However, in graphene there are also interband excitations
from the lower to the upper cone, and the associated interband single-particle continuum affects the plasmon dispersion for larger $q$.
Such interband excitations are absent in the 2DEG model.

In this paper we study a type of collective excitation that has so far not attracted much attention in graphene, namely, spin waves.
As illustrated on the right side of Fig. \ref{fig1}, we consider doped, magnetized graphene in which the spin-up and spin-down bands are split
by an effective Zeeman energy $Z^*$. The upper right-hand panel shows inter- and intraband spin-flip excitations, and the lower right-hand panel shows
the dispersion of a collective spin-flip mode or spin wave.

The corresponding spin waves in magnetic 2DEGs have been well studied theoretically and experimentally \cite{Jusserand1992,Jusserand1993,Perez2007,Perez2009,Baboux2012,Baboux2013,Perez2016,Karimi2017,DAmico2019,Maiti2015a,Maiti2015b,Maiti2016,Maiti2017,Kung2017}.
On the other hand, apart from a recent study based on Fermi-liquid theory \cite{Raines2021}, spin waves in doped graphene have not been investigated to our knowledge.
A different type of collective excitation, known as magnetoplasmon, has been more widely studied in graphene, including edges, nanoribbons, and other graphene nanostructures
\cite{Roldan2010,Pyatkovskiy2011,Balev2011,Culchac2012,Wang2012,Crassee2012,Jin2016,Sokolik2019,Jiao2021}. Magnetoplasmons occur in the presence
of Landau level quantization induced by perpendicular magnetic fields. Here, by contrast, we will consider situations where the spin splitting
can be thought of as being induced by in-plane magnetic fields, hence there are no Landau levels.

Traditional band theory, based on density-functional theory (DFT), has been extremely successful in describing materials with Dirac-like topological
features \cite{Bansil2016}. Thus, in principle, we could calculate the graphene band structure using, for instance, the local spin-density approximation (LSDA),
and then obtain the spin-wave dispersions using linear-response theory based on time-dependent density-functional theory (TDDFT) \cite{Ullrich2012}, similar to the standard way of calculating magnons in magnetic materials from first principles \cite{Savrasov1998,Buczek2009,Zakeri2014}.

On the other hand, electrons in graphene and other topological materials close to the Dirac points are well described by simple tight-binding model Hamiltonians
\cite{Neto2009,Shen}, which defines the model system of a 2D Dirac fermion gas; the purpose of this paper is to study spin waves
within this model system. This simplifies the task enormously, since a fully-fledged band-structure calculation is not needed, and the
electronic single-particle states are known analytically.

However, when it comes to the calculation of spin waves, the Dirac fermion model presents us with an interesting challenge. The formation of
spin waves in itinerant electron systems is due to electronic many-body effects beyond the RPA. Many-body effects in graphene have been
studied in the literature, see e.g. \cite{Kotov2012,Sodemann2012}. In the language of TDDFT,
these are dynamical exchange-correlation (xc) effects, which have to be approximated as functionals of the (spin) density.
Most standard approximations in DFT, such as the LSDA or gradient-corrected functionals \cite{Perdew1996}, are based on the homogeneous electron gas
(or the 2DEG \cite{Ullrich2002,Rasanen2010}); but a homogeneous electron gas is not an appropriate reference system for Dirac fermions. Thus, the standard DFT
functionals are not applicable.

A way out of this dilemma is to use xc functionals which are not tied to any reference system, such as the so-called
``orbital functionals'' of (TD)DFT \cite{Kummel2008}. In this paper we will use two orbital-dependent approximations, namely, the local
exchange functional of Slater \cite{Slater1951}, and the Singwi-Tosi-Land-Sj{\"o}lander (STLS) approach to include correlation \cite{Singwi1968,GiulianiVignale}.
These functionals were recently used to study the structure and dynamics in Hubbard systems with noncollinear magnetism \cite{Ullrich2018,Pluhar2019}.
Here, we will use them to analyze the spin-wave dispersion and spin stiffness of doped magnetized graphene as a function of doping concentration and
degree of spin polarization.

This paper is organized as follows: Section \ref{sec2} presents the necessary formal background for describing collective excitations within spin-TDDFT,
namely, linear-response theory for noncollinear spins and the definitions of the Slater and STLS approximations. In
Section \ref{sec3} we define our model for Zeeman-split Dirac fermions in graphene, and show how to calculate spin waves using the Slater and STLS approximations.
Section \ref{sec4} then presents results for spin-wave dispersions and spin stiffnesses for various parameters, and discusses prospects for experimental observation.
Conclusions are given in Section \ref{sec5}.
Further information regarding the derivation of the noninteracting response function, a discussion of the magnetic fields required to produce
spin-split bands, and additional numerical details are given in the Appendix.
Atomic units ($e=m = \hbar = 4\pi \epsilon_0 = 1$) are used throughout.

\section{Collective excitations with Spin-TDDFT}\label{sec2}

\subsection{Linear response formalism}\label{sec2A}

The excitations of interacting electronic systems are encoded in the many-body response function \cite{GiulianiVignale}.
Here, we are specifically concerned with spin waves, which are collective spin-flip modes; thus, a spin-dependent linear-response
formalism is required, which will be based on TDDFT for noncollinear spins. In this framework, the
basic variable is the spin-density-matrix,
\begin{equation}\label{spindensitymatrix}
\underline{\underline n}(\bfr) = \left( \begin{array}{cc} n_{\uu}(\bfr) & n_{\ud}(\bfr) \\ n_{\du}(\bfr) & n_{\dd}(\bfr) \end{array}\right).
\end{equation}
Alternatively, the theory can be formulated in terms of the particle density $n(\bfr) = n_{\uu}(\bfr)+n_{\dd}(\bfr)$ and the
magnetization vector $\bfm(\bfr) = \mbox{tr}\{ {\bm \sigma} \underline{\underline n}(\bfr)\}$, where $\bm\sigma$ is the vector of Pauli matrices.

Let us now consider the response of the system to a frequency-dependent perturbation $\delta\underline{\underline v}(\bfr,\omega)$, which has a similar
matrix form as the spin-density-matrix (\ref{spindensitymatrix}), and couples to the charge and spin degrees of freedom.
The linear spin-density-matrix response is given by
\begin{equation}\label{nsd}
\delta\underline{\underline n}(\bfr,\omega) = \int d\bfr' \mathbb{\bbchi}(\bfr,\bfr',\omega) \delta \underline{\underline v}(\bfr',\omega)\:,
\end{equation}
where $\mathbb{\bbchi}$ is the many-body spin-density-matrix response tensor.
In TDDFT, Eq. (\ref{nsd}) is rewritten as
\begin{equation}\label{nsdTDDFT}
\delta\underline{\underline n}(\bfr,\omega) = \int d\bfr' \mathbb{\bbchi}^{(0)}(\bfr,\bfr',\omega) \delta \underline{\underline v}^{\rm eff}(\bfr',\omega)\:,
\end{equation}
where $\mathbb{\bbchi}^{(0)}$ is the response tensor of the noninteracting Kohn-Sham system, and the effective perturbation is defined as the sum
of the physical perturbation plus a Hartree and exchange-correlation (Hxc) contribution,
$\delta \underline{\underline{v}}^{\rm eff} = \delta \underline{\underline{v}} + \delta \underline{\underline{v}}^{\rm Hxc}$. The latter is given by
\begin{equation}\label{vsd}
\delta \underline{\underline{v}}^{\rm Hxc} (\bfr,\omega) =
\int\!d\bfr' \: \mathbb{f}^{\rm Hxc}(\bfr,\bfr',\omega) \delta\underline{\underline n}(\bfr',\omega) \;.
\end{equation}
The Hartree part of the Hxc kernel is diagonal in the spin indices,
$f^{\rm H}_{\sigma\sigma',\tau \tau'}(\bfr,\bfr',\omega) = \delta_{\sigma\sigma'}\delta_{\tau \tau'} / |\bfr - \bfr'|$.
The remainder, the xc kernel $\mathbb{f}^{\rm xc}(\bfr,\bfr',\omega)$, needs to be approximated.

To obtain the excitation energies of the physical systems, we need an explicit expression for the interacting spin-density matrix response tensor $\mathbb{\bbchi}$.
Comparing the two response equations, Eqs. (\ref{nsd}) and (\ref{nsdTDDFT}), leads to
\begin{equation}\label{fullresponse}
\mathbb{\bbchi} = \Big( \mathbb{1} - \mathbb{\bbchi}^{(0)} \mathbb{f}^{\rm Hxc}\Big)^{-1}\mathbb{\bbchi}^{(0)}.
\end{equation}
The excitations are at those frequencies where $\mathbb{\bbchi}$ is not invertible. This can happen in two ways:
when $\mathbb{\bbchi}^{(0)}$ is not invertible, or when $\mathbb{1} - \mathbb{\bbchi}^{(0)} \mathbb{f}^{\rm Hxc}$ is not invertible.
The former yields the single-particle excitation spectrum, the latter the collective excitations.

Thus, all we need is the noninteracting response tensor $\mathbb{\bbchi}^{(0)}$ and an approximation for $\mathbb{f}^{\rm xc}$.
Here, we consider the special case where the ground state is such that the system is uniformly magnetized, i.e., the spins are collinear.
In that case, $\mathbb{\bbchi}^{(0)}$ only has four nonvanishing components: $\chi_{\uu,\uu}^{(0)}, \chi_{\dd,\dd}^{(0)}, \chi_{\ud,\ud}^{(0)},\chi_{\du,\du}^{(0)}$.
The xc tensor, on the other hand, has the nonvanishing elements
$f^{\rm xc}_{\uu,\uu} , f^{\rm xc}_{\dd,\dd} , f^{\rm xc}_{\uu,\dd} , f^{\rm xc}_{\dd,\uu} , f^{\rm xc}_{\ud,\ud} , f^{\rm xc}_{\du,\du}$. With this, we find that
$\mathbb{1} - \mathbb{\bbchi}^{(0)} \mathbb{f}^{\rm Hxc}$ can be represented in the following $4\times 4$ matrix form:
\begin{widetext}
\begin{equation} \label{InteractingResponseSpin}
\mathbb{1} - \mathbb{\bbchi}^{(0)} \mathbb{f}^{\rm Hxc}=\left( \begin{array}{cccc}
1-(w+f_{\ua\ua,\ua\ua}^{\rm xc})\chi_{\ua\ua,\ua\ua}^{(0)} & 0 & 0 & -(w +f_{\ua\ua,\da\da}^{\rm xc})\chi_{\ua\ua,\ua\ua}^{(0)}\\
0 & 1-f_{\ua\da,\ua\da}^{\rm xc}\chi_{\ua\da,\ua\da}^{(0)} & 0 & 0 \\
0 & 0 & 1-f_{\da\ua,\da\ua}^{\rm xc}\chi_{\da\ua,\da\ua}^{(0)} &  0\\
-(w+f_{\da\da,\ua\ua}^{\rm xc})\chi_{\da\da,\da\da}^{(0)} & 0 &0  & 1-(w+f_{\da\da,\da\da}^{\rm xc})\chi_{\da\da,\da\da}^{(0)}\\
  \end{array}\right),
\end{equation}
\end{widetext}
where $w$ represents the Coulomb interaction. This matrix is noninvertible if its determinant is zero.
We can rearrange the matrix in block diagonal form, so that the determinant factors into a product of the determinants of two $2\times 2$ matrices:
\begin{equation}
\mbox{det}|\mathbb{1} - \mathbb{\bbchi}^{(0)} \mathbb{f}^{\rm Hxc}| = \mbox{det}|\underline{\underline{M}}_L| \; \mbox{det}|\underline{\underline{M}}_T| \:,
\end{equation}
where the longitudinal block is
\begin{equation}
\underline{\underline{M}}_L = \! \left( \begin{array}{cc} 1-(w+f_{\ua\ua,\ua\ua}^{\rm xc})\chi_{\ua\ua,\ua\ua}^{(0)} & -(w +f_{\ua\ua,\da\da}^{\rm xc})\chi_{\ua\ua,\ua\ua}^{(0)}\\
-(w+f_{\da\da,\ua\ua}^{\rm xc})\chi_{\da\da,\da\da}^{(0)}  & 1-(w+f_{\da\da,\da\da}^{\rm xc})\chi_{\da\da,\da\da}^{(0)}\end{array}\right)
\label{eq:MLongitudinal}
\end{equation}
and the transverse block is
\begin{equation}
\underline{\underline{M}}_T = \left( \begin{array}{cc} 1-f_{\ua\da,\ua\da}^{\rm xc}\chi_{\ua\da,\ua\da}^{(0)} &0\\
0 & 1-f_{\da\ua,\da\ua}^{\rm xc}\chi_{\da\ua,\da\ua}^{(0)}  \end{array}\right).
\label{eq:MTransverse}
\end{equation}
Here, longitudinal and transverse refers to the spin quantization axis. Thus, the condition $\mbox{det}|\underline{\underline{M}}_L|=0$ yields
the longitudinal (or spin-conserving) collective excitations, and $\mbox{det}|\underline{\underline{M}}_T|=0$ yields the transverse (or spin-flip) collective excitations.
The former are the usual plasmon mode and a longitudinal spin excitation, and the latter are the spin waves.
In the following we will make the adiabatic approximation for the xc kernel, i.e., we ignore the frequency dependence of the $f^{\rm xc}$'s.

\subsection{Construction of dispersion relations}\label{sec2B}

We see that both longitudinal unpolarized charge (plasmon) and transverse spin (spin-wave) collective excitations satisfy equations of the schematic form
\begin{equation}\label{eq_coll}
  1-f(q) \chi(q,\w) = 0,
\end{equation}
where $f$ and $\chi$ are placeholders for the corresponding functions specific to the underlying system.

In order to obtain the dispersion relation, $\w(q)$, that satisfies this condition, one must be able to invert the function $\chi(q,\w)$ for $\w$.
It is possible to approximate the inverse of $\chi$ to arbitrary degree using a technique called series reversion \cite{Abramowitz}.
The procedure consists of obtaining a series approximation of $\chi$ in $\w$ to arbitrary degree, obtain the inverse of the series using series reversion, and finally evaluate the inverse series at $1/f(q)$.

We expand the response in $\w$ around $\omega(q=0)\equiv \w_0$:
\begin{equation}
  X_n[q,\w_0](\w) = \sum_{l=0}^n \frac{\chi^{(0,l)}(q,\w_0)}{l!}  (\w-\w_0)^l,
\end{equation}
where $X_n[q,\w_0](\w)$ is the truncated series approximation of $\chi(q,\w)$, and we define a shorthand for the partial derivatives of a function of the form $g(x,y)$:
\begin{equation}
  g^{(m,n)} (a,b) = \left( \frac{\partial^m}{\partial x^m}\frac{\partial^n}{\partial y^n} g(x,y) \right)_{x=a,y=b}.
\end{equation}
We then use $X$ in place of $\chi$ in  Eq. (\ref{eq_coll}),
\begin{equation}
  \chi(q,\w) \approx X_n [q,\w_0](\w)=\frac{1}{f(q)} ,
\end{equation}
and inversion of this yields
\begin{equation}
  \w(q) \approx X_n^{\text{inv}}[q,\w_0]\left(\frac{1}{f(q)}\right).
\end{equation}
The details of this inverse series are left to the specifics of the system.

\subsection{The Slater and STLS approximations}\label{sec2C}

To calculate spin-wave excitations, the transverse xc kernels $f^{\rm xc}_{\ua\da,\ua\da},f^{\rm xc}_{\da\ua,\da\ua}$ are needed.
As discussed in the Introduction, we shall work with two orbital-dependent approximations:
Slater and STLS. Both xc kernels are independent of the frequency $\omega$.

The Slater exchange kernel \cite{Petersilka1996,GiulianiVignale} was generalized
for noncollinear spin dynamics in Ref. \cite{Ullrich2018}. In the limit where the ground state has collinear spin, the Slater exchange kernel has the following elements:
\begin{eqnarray}\label{eq:Slater1}
f^{\rm x,S}_{\sigma\sigma,\sigma\sigma}(\bfr,\bfr') &=& -\frac{|\gamma_{\sigma\sigma}(\bfr,\bfr)|^2}{n_{\sigma\sigma}(\bfr)n_{\sigma\sigma}(\bfr')|\bfr - \bfr'|}
\\
f^{\rm x,S}_{\sigma\bar\sigma,\sigma\bar\sigma}(\bfr,\bfr') &=& -4\frac{\gamma_{\sigma\sigma}(\bfr,\bfr')\gamma_{\bar\sigma\bar\sigma}(\bfr',\bfr')}{n(\bfr)n(\bfr')|\bfr - \bfr'|} \:,
\label{eq:Slater2}
\end{eqnarray}
where $\gamma_{\sigma\sigma'}(\bfr,\bfr')$ is the spin-resolved reduced one-body density matrix of the Kohn-Sham system.

For an unpolarized 2DEG, the Fourier transform of the Slater exchange kernel was worked out in Ref. \cite{Karimi2014}.
The generalization to the spin-polarized 2DEG, with spin densities $n_\sigma$ and total density $n = n_\ua+n_\da$
and corresponding Fermi wavevectors  $k_{F\sigma} = \sqrt{2\pi n_\sigma}$ and $k_F = \sqrt{2\pi n}$, is as follows:
\begin{eqnarray}
f^{\rm x,S}_{\sigma\sigma,\sigma\sigma}(q) &=& -\frac{8\pi}{k_{F\sigma}^2} \int_0^\infty \frac{dx}{x^2} J_0(qx)J_1^2(k_{F\sigma}x)\\
f^{\rm x,S}_{\sigma\bar\sigma,\sigma\bar\sigma}(q) &=& -\frac{8\pi k_{F\sigma}k_{F\bar\sigma}}{k_{F}^4} \int_0^\infty \frac{dx}{x^2} J_0(qx)J_1(k_{F\sigma}x)J_1(k_{F\bar\sigma}x)
\nonumber\\
&&
\end{eqnarray}
where $J_0$ and $J_1$ denote Bessel functions of the first kind.
The corresponding expressions for graphene will be discussed below in Sec. \ref{sec3D}.

The STLS xc kernel \cite{Singwi1968,GiulianiVignale} was generalized to systems
with noncollinear spin in Ref. \cite{Ullrich2018}. More precisely, the noncollinear formulation of Ref. \cite{Ullrich2018} involved a simplified, scalar form
of the original STLS scheme, termed sSTLS. Again, let us consider the special case where the ground state is collinear.
The elements of the xc tensor $\mathbb{f}^{\rm xc,sSTLS}(\bfr,\bfr')$ are then given by
\begin{eqnarray}\label{sSTLSkernel}
f^{\rm xc,sSTLS}_{\sigma\sigma',\alpha\alpha'}(\bfr,\bfr')
&=&
\frac{1}{|\bfr - \bfr'|} {\cal R}_{\sigma\sigma',\alpha\alpha'}(\bfr,\bfr')\\
&\times&\left[
\SC_{\sigma\sigma',\alpha\alpha'}(\bfr,\bfr') - \delta_{\sigma\alpha}\delta(\bfr-\bfr') n_{\alpha'\sigma'}(\bfr)\right], \nonumber
\end{eqnarray}
where ${\cal R}_{\sigma\sigma',\alpha\alpha'}$ are the elements of the following matrix:
\begin{equation}
\mathbb{R} = \left( \begin{array}{cccc}
\frac{1}{n_\ua(\bfr) n_\ua(\bfr')} & \frac{2}{n(\bfr) n_\ua(\bfr')} & \frac{2}{n(\bfr) n_\ua(\bfr')}& \frac{1}{n_\da(\bfr) n_\ua(\bfr')}\\[2mm]
\frac{2}{n_\ua(\bfr) n(\bfr')}     & \frac{4}{n(\bfr) n(\bfr')}     & \frac{4}{n(\bfr) n(\bfr')}    & \frac{2}{n_\da(\bfr) n(\bfr')}\\[2mm]
\frac{2}{n_\ua(\bfr) n(\bfr')}     & \frac{4}{n(\bfr) n(\bfr')}     & \frac{4}{n(\bfr) n(\bfr')}    & \frac{2}{n_\da(\bfr) n(\bfr')}\\[2mm]
\frac{1}{n_\ua(\bfr) n_\da(\bfr')} & \frac{2}{n(\bfr) n_\da(\bfr')} & \frac{2}{n(\bfr) n_\da(\bfr')}& \frac{1}{n_\da(\bfr) n_\da(\bfr')}
\end{array}\right).
\end{equation}
$\SC_{\sigma\sigma',\alpha\alpha'}(\bfr,\bfr')$ are the elements of the generalized static structure factor:
\begin{equation}\label{structurefactor}
\mathbb{S}(\bfr,\bfr') = -\frac{1}{\pi}\int_0^\infty \Im \mathbb{\bbchi}(\bfr,\bfr',\omega) d\omega \:.
\end{equation}
The response tensor $\mathbb{\bbchi}$, in turn, follows from Eq. (\ref{fullresponse}) evaluated with
the sSTLS kernel (\ref{sSTLSkernel}), which closes the self-consistency loop. If in the first step of the iteration  Eq. (\ref{structurefactor}) is initialized
with $\mathbb{\bbchi}^{(0)}$, then Eq. (\ref{sSTLSkernel}) yields the Slater exchange kernel. Correlation enters in the subsequent iteration steps.

Consider again the homogeneous 2D case and carry out a Fourier transformation of Eq. (\ref{sSTLSkernel});
specifically, for the transverse xc kernel in sSTLS approximation we obtain
\begin{equation} \label{eq:fxcspinSTLSscalar}
f^{\rm xc, sSTLS}_{\sigma\bar\sigma,\sigma \bar\sigma}(q) = \frac{4}{n^2} \sum_{\bfq'} \: v_{q'}
[\SC_{\sigma\bar\sigma,\sigma \bar\sigma}(\bfq - \bfq') - n_{\bar\sigma\bar\sigma}],
\end{equation}
where $v_q=2\pi/q$.
We also introduce an alternative form of the xc kernel, which directly generalizes the original STLS approach \cite{Singwi1968}:
\begin{equation} \label{eq:fxcspinSTLSforce}
f^{\rm xc, STLS}_{\sigma\bar\sigma,\sigma\bar\sigma}(q) = \frac{4}{n^2}\sum_{\bfq'}  \:\frac{ \bfq \cdot \bfq'}{q^2} v_{q'}
[\SC_{\sigma\bar\sigma,\sigma\bar\sigma}(\bfq - \bfq') - n_{\bar\sigma\bar\sigma}] \:.
\end{equation}
The difference between the two schemes is that the scalar sSTLS is based on the effective potential whereas the
full STLS is based on the effective force \cite{GiulianiVignale}. Expressing the full STLS kernel in real space and for inhomogeneous systems causes some technical complications,
since forces formally couple to currents rather than densities \cite{Dobson2009}. However, in the homogeneous case the transition from Eq. (\ref{eq:fxcspinSTLSscalar})
to (\ref{eq:fxcspinSTLSforce}) is straightforward.

For graphene, the construction of the (s)STLS xc kernels involves some subtleties, which we will discuss below.


\section{Collective excitations in graphene}\label{sec3}

\subsection{Model: Dirac fermions with Zeeman splitting}\label{sec3A}

The tight-binding model is commonly used to describe the band structure of graphene \cite{Neto2009}.
We consider a generalization in which the spin-up and spin-down bands are split by a Zeeman energy $Z^*$.
Thus, we consider a Hamiltonian of the form
\begin{equation}\label{grapheneTB}
\hat H = -t \sum\limits_{\left<l,m\right>,\si} (\hat c_{l\si}^\dagger \hat c_{m\si} + \text{H.c.} ) + \frac{Z^*}{2} \sum\limits_{j,\si} s_\si
\hat c_{j\si}^\dagger \hat c_{j\si} ,
\end{equation}
where $t$ $(\approx 2.8$ eV) is the nearest-neighbor hopping energy, $\hat c_{l\si}^\dagger (\hat c_{l\si})$ is the creation (annihilation) operator for an
electron with spin $\si$ at the $l$th site, and $s_{\si} = +1$ and $-1$ for $\si=\uparrow$ and $\downarrow$, respectively.
The sum over $\left<l,m\right>$ is restricted to nearest neighboring sites.

In our model, $Z^*$ is treated as an adjustable parameter. Notice that $Z^*$ denotes the Zeeman energy renormalized by electronic many-body effects, which in
general differs from the bare Zeeman energy $Z$ \cite{DAmico2019}. To determine $Z^*$ from first principles would require a self-consistent calculation
of the band structure in the presence of an externally applied uniform in-plane magnetic field, or a magnetic field induced by proximity to a ferromagnetic substrate;
how this could be realized will be further discussed in Sec. \ref{sec4B}. In Appendix \ref{appB} we calculate the effective magnetic field strengths
required to produce given values of $Z^*$ in graphene.

\begin{figure}
  \begin{center}
    \includegraphics[width=\linewidth]{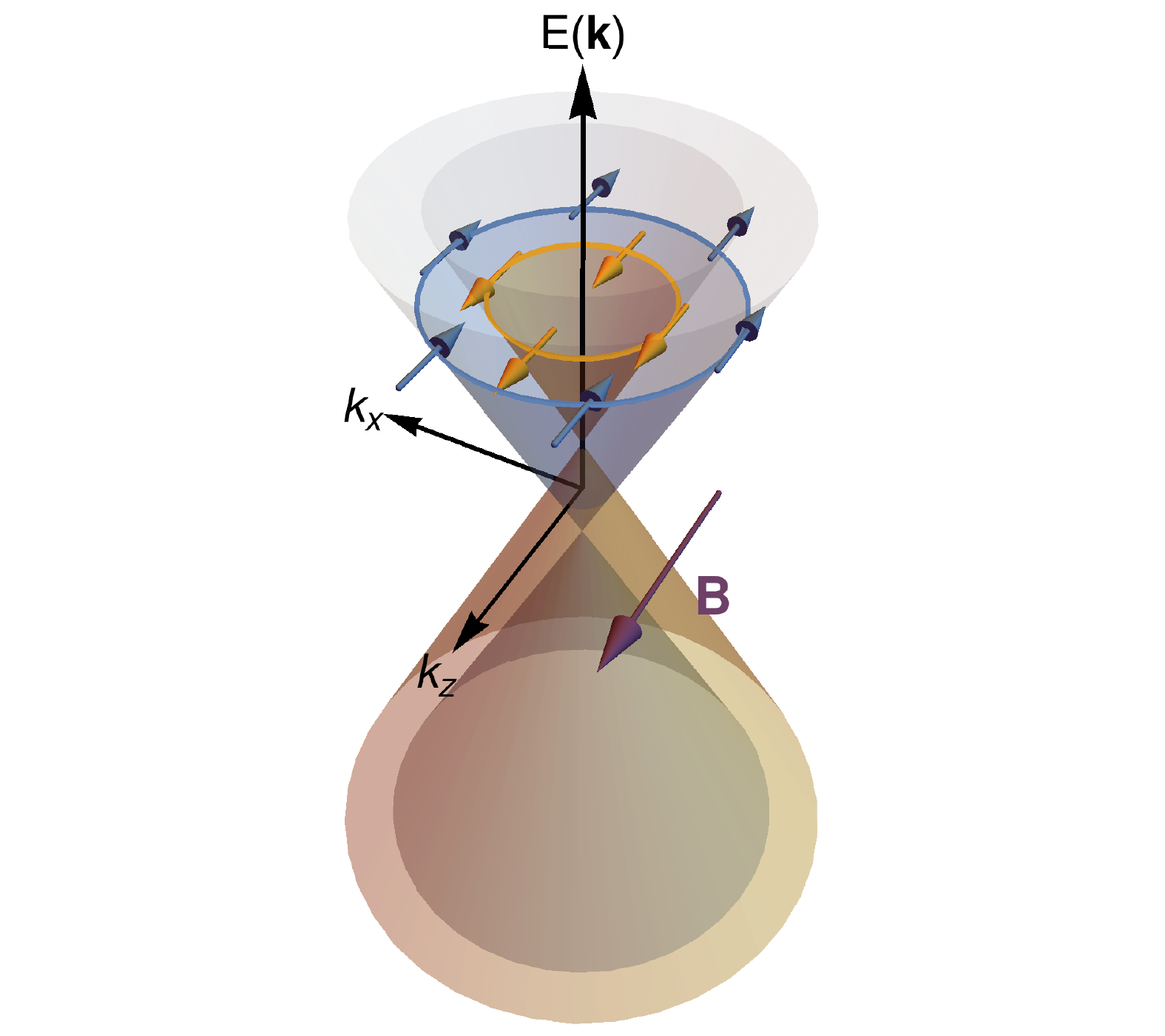}
    \caption{\small Spin-split Dirac cones of doped graphene. The Zeeman splitting gives rise to
    different Fermi surfaces for the majority and minority spins.}
    \label{fig:GrapheneStructure}
  \end{center}
\end{figure}

We are particularly interested in the electronic states close to the Dirac points. Instead of working with
the eigenstates of the full tight-binding Hamiltonian (\ref{grapheneTB}), one can consider the
eigenstates of the reduced Hamiltonian $\hat H_K$, valid around the $K$ point of the graphene Brillouin zone \cite{Neto2009,Ando2006}:
\begin{equation}\label{grapheneKK}
\hat H_{K} =\gamma \left(\hat \si_x^{(b)} k_x + \hat \si_y^{(b)} k_y\right) + \frac{Z^*}{2} \hat \si^{(s)}_z .
\end{equation}
Here, $\gamma = 3at/2$, where $a$ is the C-C bond length in graphene, $\hat \sigma_{x,y,z}$  are Pauli matrices operating on the
band $(b)$ or spin $(s)$ degrees of freedom, and $\bfk = (k_x,k_y)$ is a wave vector measured with respect to the Dirac point $K$.
The energy eigenvalues of $\hat H_K$ are
\begin{equation}\label{eq:GrapheneEnergySpin}
  \varepsilon_{\bfk b\si} = b\gamma |\bfk| + s_\si \frac{Z^*}{2},
\end{equation}
where $b=\pm 1$ is the band index. The spin-split Dirac cones are illustrated in Fig. \ref{fig:GrapheneStructure}.
The associated eigenstates are
\begin{equation}\label{psiK}
  \psi_{\bfk b\si}^K (\bfr) = \frac{e^{i \bfk \cdot \bfr}}{\sqrt{2}} \left( \begin{array}{c}
    e^{-i \phi_k} \\ b
  \end{array} \right)  \otimes \bfs_\si \:,
\end{equation}
involving the product of a two-component pseudospinor (since there are two sites within a unit cell)
with the two-component (up/down) spinor $\bfs_\si$. The eigenstates $\psi_{\bfk b\si}^{K'} (\bfr)$
around the $K'$ point are obtained by replacing $b$ with $-b$ in Eq. (\ref{psiK}).

\subsection{Noninteracting response function}\label{sec3B}

As discussed in Sec. \ref{sec2A}, the noninteracting response tensor $\mathbb{\bbchi}^{(0)}$ is the fundamental object needed to calculate collective excitations.
The generic definition is
\begin{equation}\label{eq:KSResponse}
  \mathbb{\bbchi}^{(0)} (\bfr,\bfrp,\w) = \sum\limits_{jl} (f_l-f_j) \frac{\psi_j(\bfr) \psi_l^\dagger(\bfr) \psi_j^\dagger(\bfrp) \psi_l(\bfrp)}
  {\w - (\varepsilon_j - \varepsilon_l) +i\eta} \:,
\end{equation}
where the $\psi_j$ are single-particle spinor wave functions labeled by a set of quantum numbers $j$, $\varepsilon_j$ are the associated single-particle energies,
$f_j$ are occupation numbers (here, either 0 or 1), and $\eta$ is a positive infinitesimal.

The noninteracting response function of graphene (within the Dirac fermion model) is obtained by substituting the single-particle
energies (\ref{eq:GrapheneEnergySpin}) and eigenstates (\ref{psiK}) into Eq. (\ref{eq:KSResponse}). The spin-independent form of the graphene response function is
well known from the literature \cite{Ando2006,Wunsch2006,Abedinpour2011}; here, we generalize it to the spin-dependent case.
Details of the derivation are given in Appendix \ref{appA}.
Furthermore, instead of real frequencies $\omega$ we evaluate the response function for complex frequencies $z$, which
has certain technical advantages, as discussed in Appendix \ref{appC}.
The result for the non-spin-dependent response function at $Z^*=0$, $\chi^{(0)} = \chi^{(0)}_{\uu,\uu} + \chi^{(0)}_{\dd,\dd}$,
is as follows:
\begin{eqnarray}\label{eq:NIResponse}
\frac{\chi^{(0)}(q,z)}{g_s g_v} &=&
-\frac{\kf}{2\pi\gamma} -\frac{q}{16 \gamma \sqrt{1-\left(\frac{z}{\gamma q}\right)^2}} \nonumber\\
&+&
\sum_{\alpha  }^{\pm 1}\frac{ q }{16\pi\gamma }  G\left(\frac{\alpha z}{\gamma q},\frac{\alpha z +2 \gamma \kf}{\gamma q}\right),
\end{eqnarray}
where $g_s$ and $g_v$ are the spin and valley degeneracies, respectively,
$\kf=\sqrt{4\pi n/g_s g_v}$ is the Fermi wave vector associated with the 2D electron density of the upper bands $n$,
and the function $G(a,x)$ is defined as
\begin{equation}\label{eq:ResponseG}
  G(a,x) = \frac{x(x^2-1)-\sqrt{1-x^2}\arcsin x}{(ax-1)\sqrt{\frac{(1-x^2)(1-a^2)}{(1-ax)^2}}} .
\end{equation}
Eq. (\ref{eq:NIResponse}) reduces to the real frequency form from previous work by taking $\lim_{\eta \to 0^+} (z= \w + i\eta)$.
It is important to note that Eq. (\ref{eq:ResponseG}) cannot be further reduced since the proper branch cuts must be preserved.

The transverse spin-dependent response functions at finite $Z^*$,  $\chi^{(0)}_{\ud,\ud}$ and  $\chi^{(0)}_{\du,\du}$, are given by
\begin{eqnarray}
\frac{\chi^{(0)}_{\s\bar\s,\s\bar\s}(q,z)}{g_v}
&=&
-\frac{\ks+\ksb}{4\pi \gamma}
-\frac{q}{16\gamma\sqrt{1-\left(\frac{z+\essb}{\gamma q}\right)^2}} \nonumber\\
&+&
\frac{q}{16\pi \gamma} \left[G\left( \frac{z+\essb}{\gamma q}, \frac{ z+\essb +2\gamma \ks}{\gamma q}\right)\right. \nonumber\\
&& \left. -G\left( \frac{z+\essb}{\gamma q}, \frac{z+\essb -2\gamma \ksb}{\gamma q}\right)\right],
  \label{eq:NIResponseSpin}
\end{eqnarray}
where $\bar\sigma = \da$ if $\sigma=\ua$ and vice versa, $\ks = \sqrt{2\pi n_\sigma}$ is the Fermi wavevector associated with the spin density of the upper band $n_\sigma$,
and $\essb = \varepsilon_{\bfk b \si} - \varepsilon_{\bfk b \sib} = \frac{s_\si-s_\sib}{2}Z^*$ is the single-particle spin-flip energy, which is independent of $b$ and $\bfk$
for the simple Zeeman splitting considered here.

\begin{figure}
  \includegraphics[width=\linewidth]{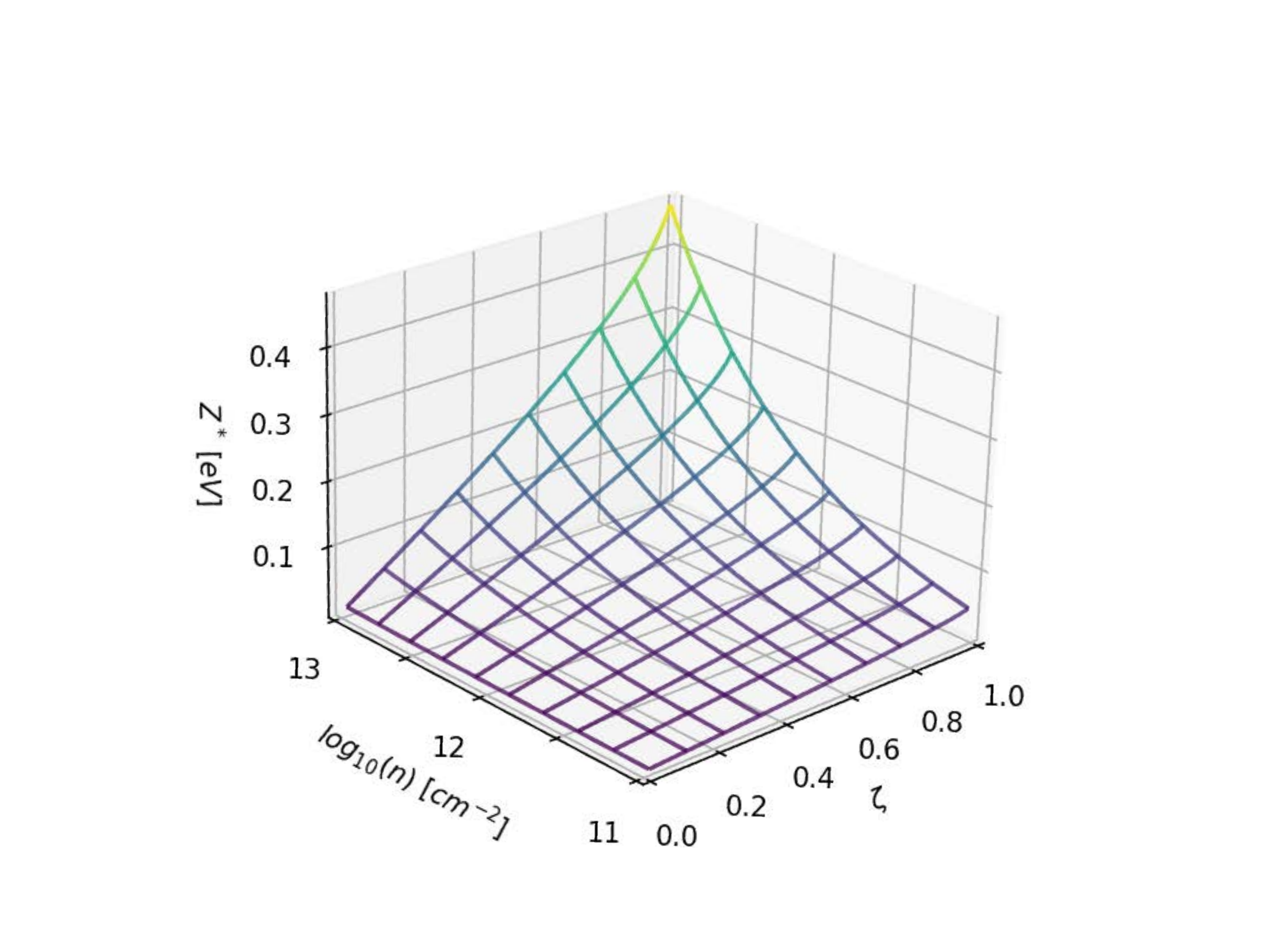}
  \caption{The renormalized Zeeman energy, $Z^*$, as a function of doping concentration $n$ and spin polarization $\zeta$.}
  \label{fig:Zstar}
\end{figure}

The renormalized Zeeman energy can be expressed as
\begin{equation}
  Z^* = \gamma |\ku-\kd| = \gamma \sqrt{\pi n} \left|\sqrt{1+\z}-\sqrt{1-\z} \right| \: ,
  \label{eq:Zstar}
\end{equation}
where $\zeta = (n_\ua - n_\da)/n$ is the spin polarization of the conduction electrons.
Figure \ref{fig:Zstar} shows $Z^*$ within the density-polarization parameter space.
The range of doping densities $n$ ($10^{11}-10^{13}\text{cm}^{-2}$) is chosen such that the Dirac model is still valid, i.e.,
the Fermi level does not reach those parts of the conduction band where the band dispersion deviates significantly from linearity.
We find that $Z^*$ can reach values of a few hundreds of meV for strong doping and high degrees of spin polarization.

\subsection{Mode dispersions and spin stiffness}\label{sec3C}

\subsubsection{Plasmons}

Let us first consider the spin-unpolarized case.
The graphene plasmon dispersion energy goes to zero as $q$ approaches zero, see Fig. \ref{fig1}.
This is problematic for the inverse series procedure because the response function has a singularity in the $q$-$\w$ plane at $(0,0)$.
The limit of the response function depends on the direction as $(0,0)$ is approached.
For collective excitations, it is important to calculate the dispersion in the dynamical long-wavelength limit (DLWL) \cite{GiulianiVignale}, i.e. $\w \gg v_F q$,
where $v_F$ is the Fermi velocity. It is useful to introduce the parameter
$\nu = \w/\gamma q$, which defines the slope of a line passing through the origin and thus the direction of the limit.
In order to obtain the low-$q$ behavior and still be in the DLWL, we expand the response function in $\nu$ near infinity.
This is equivalent to expanding in $1/\nu$ near 0.

The first few terms of the series expansion of the response function of Eq. (\ref{eq:NIResponse}) in $\nu$ are
\begin{equation}
  \chi^{(0)}(q,\gamma q \nu) =
  \frac{ \kf}{ \pi  \gamma  \nu^2}
  +\frac{ \kf}{2 \pi  \gamma  \nu^4}
  +\bigO\left(\frac{1}{\nu^6}\right) \: .
  \label{eq:NIResponseSeries}
\end{equation}
The corresponding inverse series is
\begin{equation}
  \nu (y) = \frac{\w(y)}{\gamma q} \approx
  \sqrt{\frac{\gamma \kf}{\pi }} \left(
  \frac{1}{\gamma \sqrt{y}}
  +
  \frac{\pi}{4 \kf} \sqrt{y}
  \right)
  +\bigO\left(y^{3/2}\right) \: ,
\end{equation}
where $y=1/f^{\rm Hxc}(q)$. Thus, the plasmon dispersion relation becomes
\begin{eqnarray}
  \w_{\text{pl}}(q) &=&
  \sqrt{\frac{\gamma \kf /\pi}{f^{\rm xc}(q)+\frac{2 \pi}{q} }}
  \left( 2 \pi  + q\left( f^{\rm xc}(q)+\frac{\gamma\pi}{4\kf}  \right)\right) \nonumber\\
  &+&
  \bigO\left(\frac{1}{f^{\rm Hxc}(q)}\right)^{3/2} .
\end{eqnarray}
In principle, it is straightforward to obtain higher order terms by including more terms in Eq. (\ref{eq:NIResponseSeries}).
However, it is best to stop at the 4th order terms because of the DLWL.
The series diverges quickly for higher order terms.

\subsubsection{Spin waves}

The spin polarization $\zeta$ of the conduction electrons can be positive or negative. Let us consider the case where $\zeta>0$:  this implies
$n_\ua>n_\da$ and therefore, from Eq. (\ref{eq:GrapheneEnergySpin}), the upper (lower) of the two spin-split conduction bands has
spin $\sigma = \da$ ($\bar \sigma = \ua$). Hence, $(s_\si - s_\sib)/2 = -1$. The spin waves are obtained from Eq. (\ref{eq:MTransverse}),
but only the condition $1 - f^{\rm xc}_{\da\ua,\da\ua} \chi^{(0)}_{\da\ua,\da\ua} = 0$ is needed. The case $\zeta<0$ works in an analogous fashion, except that
$\sigma$ and $\bar \sigma$ are reversed.

The spin waves in graphene have a finite dispersion energy as $q$ goes to 0, see Fig. \ref{fig1}.
The series can be calculated directly with $\w$. The low-$q$ spin-wave dispersion relation becomes:
\begin{eqnarray}\label{SWdispersion}
\omega_{\rm sw}(q) &=& \essb \left(-1+\frac{f^{\rm xc}_0}{2\pi\gamma}(\ks+\ksb)\right) \nonumber\\
&+&
q\frac{\kf^2\z(s_\si - s_\sib){f^{\rm xc}_0}'}{2\pi} \nonumber\\
&+& \frac{q^2}{2}\frac{(s_\si - s_\sib)}{2}  S + \bigO(q^3),
\end{eqnarray}
where
\begin{eqnarray}
S &=& \frac{\gamma(\sqrt{1+s_\si \z}+\sqrt{1+s_\sib \z})}{2\kf\z}
+ \frac{\pi\gamma^2}{\kf^2 \z f^{\rm xc}_0}
\nonumber\\
&+&
\frac{(s_\si - s_\sib)}{2}\frac{f^{\rm xc}_0}{4\pi}\left(
      \ln \left| f^{\rm xc}_0-\frac{4\pi\gamma\sqrt{1+s_\sib \z}}{\kf\z(s_\si - s_\sib)} \right| \right.\nonumber\\
&-&
\left.\ln \left| f^{\rm xc}_0+\frac{4\pi\gamma\sqrt{1+s_\si \z}}{\kf\z(s_\si - s_\sib)} \right|\right)
+ \frac{\kf^2\z {f^{\rm xc}_0}''}{2\pi} .
\end{eqnarray}
Here, we use the abbreviations $f^{\rm xc}_0$, ${f^{\rm xc}_0}'$ and ${f^{\rm xc}_0}''$
for the $q=0$ limit of $f^{\rm xc}_{\si \sib,\si \sib}(q)$ and its first and second derivatives with respect to $q$, respectively.
Notice that for the Slater and STLS approximation we consider here, we have ${f^{\rm xc}_0}'=0$ and the linear term
in the spin-wave dispersion (\ref{SWdispersion}) vanishes.

The generic form of spin-wave dispersions in itinerant paramagnetic electron liquids for small $q$ is as follows:
\begin{equation}
  \w_{\rm sw}(q) = \omega_L +\frac{S_{\rm sw}}{2} q^2 +\bigO\left(q^4\right).
\end{equation}
Here, $\omega_L$ is the Larmor frequency, which indicates a collective precessional motion of all spins about the magnetic field direction.
For the case of graphene we find $\omega_L = Z^*(1 + f_0^{\rm xc}/2\gamma^2)$, which is smaller than $Z^*$ since $f_0^{\rm xc}<0$.
From electronic many-body theory one would have expected that $\omega_L = Z$ (Larmor's theorem), where $Z$ is the bare Zeeman energy, i.e., all many-body effects
cancel out exactly in the Larmor precessional state \cite{DAmico2019}. However, Larmor's theorem does not apply here since the band structure is
obtained from a tight-binding Dirac fermion Hamiltonian and not from first principles; in other words, $Z^*$ is given but $Z$ remains unknown.
One should therefore refer to $\omega_L$ more appropriately as pseudo-Larmor frequency.

The spin-wave stiffness, $S_{\rm sw}$, determines the curvature of the spin-wave dispersion
for small $q$; it can have positive or negative values depending on the parameters characterizing the electron liquid.
Here, we have  $S_{\rm sw} = (s_\si - s_\sib)S/2$.

\begin{figure}
  \includegraphics[width=\linewidth]{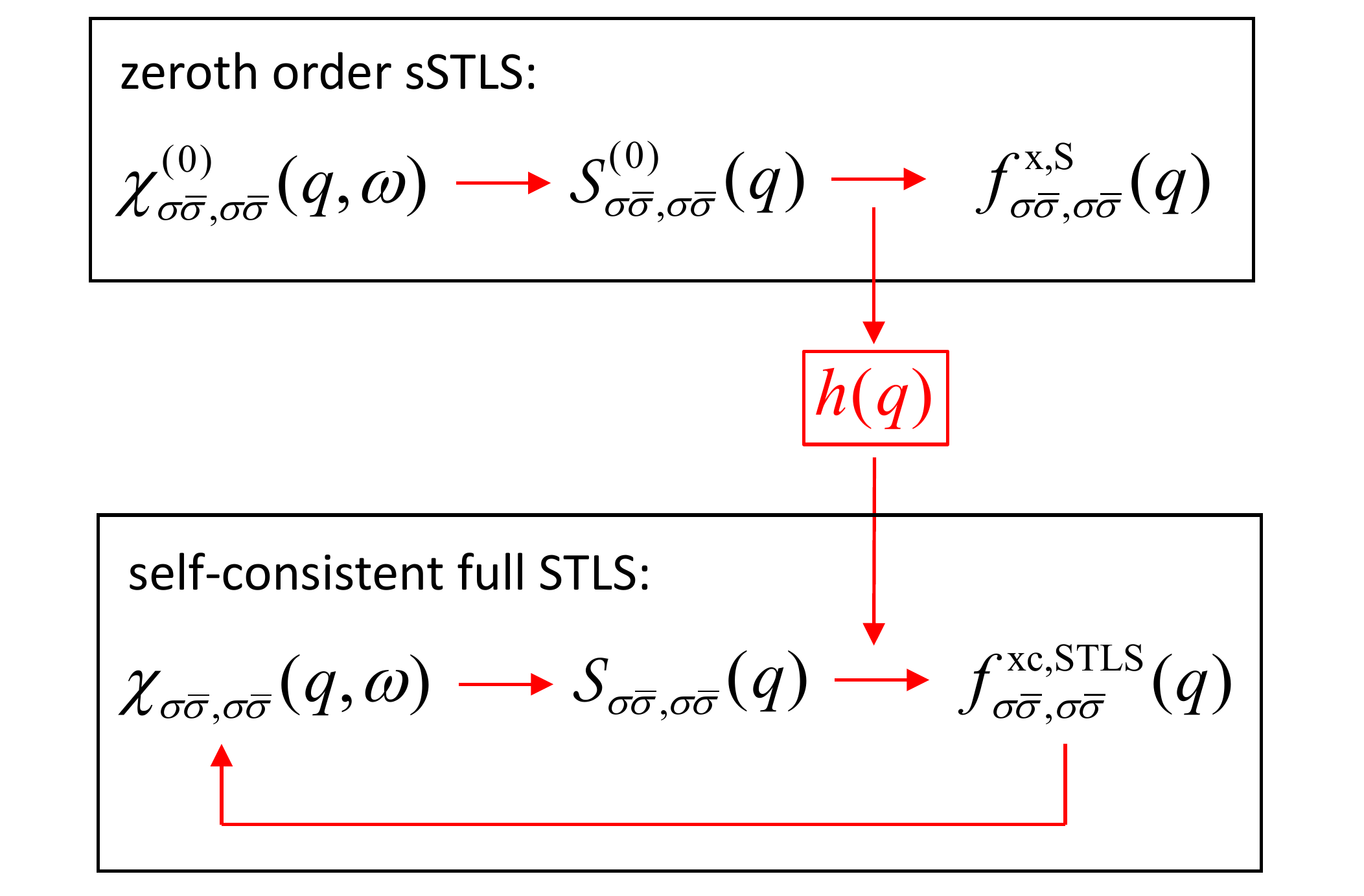}
  \caption{Modified STLS approach for Dirac fermions. To obtain the STLS xc kernel, an integration cutoff $h(q)$ is needed,
  which follows from the requirement that the zeroth iteration of the sSTLS scheme yields the Slater exchange kernel.}
  \label{fig:STLSscheme}
\end{figure}

\subsection{Slater and STLS kernels for Dirac fermions}\label{sec3D}

The Slater approximation for Dirac fermions uses the same expressions as for the 2DEG, Eqs. (\ref{eq:Slater1}) and (\ref{eq:Slater2}).
We use the graphene eigenstates (\ref{psiK}) to construct the density matrix:
\begin{equation}
\gamma_{\sigma\sigma}(\bfr,\bfr')
=
\sum_{b\bfk}^{\rm occ} \psi_{b \bfk \sigma}^\dagger(\bfrp)\psi_{b \bfk \sigma}(\bfr)
=
2\sum_{b\bfk}^{\rm occ}e^{i\bfk\cdot(\bfr-\bfrp)},
\end{equation}
where the factor 2 accounts for the valley degeneracy. Within the Dirac model, the so defined density matrix
nominally involves a diverging integral over an infinite lower band. To avoid this problem,
we impose a finite cutoff to the lower band at a wavevector $k_v$.
The natural choice for this cutoff is that which reproduces the undoped density of graphene, $n_v= 1.91 \times 10^{15}\: \rm cm^{-2}$:
\begin{equation}\label{cutoff}
  k_v = \sqrt{\pi n_v} = 0.41 \  a_0^{-1}.
\end{equation}
Since $\gamma_{\sigma\sigma}(\bfr,\bfr')$ only depends on the coordinate difference, we can make the substitution $\bfr-\bfrp = \bfrho$,
and we also define an occupation function $f_b(k)$ which depends on the band index $b$:
\begin{eqnarray}
\gamma_{\sigma\sigma}(\bfrho)
&=&
\frac{2}{(2\pi)^2}\sum_{b}\int_0^\infty k dk f_b(k) \int_0^{2\pi}d\theta e^{i k \rho \cos(\theta)} \nonumber\\
&=&
 \frac{1}{\pi}\sum_b\int_0^\infty k dk f_b(k)J_0(k\rho) \nonumber\\
&=& \frac{1}{\pi\rho}[\ks J_1(\ks \rho)+k_vJ_1(k_v \rho)].
\end{eqnarray}
With this, the transverse Slater kernel for Dirac fermions becomes:
\begin{eqnarray}
f^{\rm x,S}_{\sigma\bar\sigma,\sigma\bar\sigma}(\bfrho) &=& \frac{-4}{\pi^2 n^2 \rho^3}[\ks J_1(\ks \rho)+k_vJ_1(k_v \rho)]\nonumber\\
&\times&
[\ksb J_1(\ksb \rho)+k_vJ_1(k_v \rho)] \: .
	  \label{eq:SlaterKernelDiracR}
\end{eqnarray}
Fourier transform of this yields
\begin{eqnarray}
f^{\rm x,S}_{\sigma\bar\sigma,\sigma\bar\sigma}(q)
&=&
-\frac{8}{\pi n^2}\int_0^\infty \frac{d\rho}{\rho^2}  [\ku J_1(\ku\rho)+k_v J_1(k_v\rho)]\nonumber\\
&\times&
[\kd J_1(\kd\rho)+k_v J_1(k_v\rho)]J_0(q\rho).
\end{eqnarray}
The Slater kernel is typically dominated by the valence band contribution because of the larger number of particles compared to the conduction band.

Let us now discuss how to implement the STLS scheme for Dirac fermions.
As for the Slater approximation, a cutoff for the lower Dirac cone is necessary; otherwise, the static structure factor $\SC(q)$ diverges as $q\rightarrow\infty$
(this  happens because the structure factor is proportional to the density in the high-$q$ limit).
We introduce the same cutoff as above, Eq. (\ref{cutoff}), which ensures that the static structure factor remains finite and bounded for all $q$.

The next problem arises from the shape of the structure factor itself.
The calculation of the xc kernel converges only when $\SC(q)-n$ is asymptotically smaller than $1/q$.
However, the tail of the structure factor goes as $n_v/2 + \frac{2\kf^3-k_v^3}{6\pi q}$, which
approaches the wrong value for the density as $1/q$.
This produces an unavoidable singularity in the integrand of the xc kernel.
As a remedy, we alter the integration limits in Eq. (\ref{eq:fxcspinSTLSscalar}) such that the sSTLS xc kernel remains finite for all $q$.
We fix this limit by enforcing that the zeroth sSTLS iteration reproduces the Slater kernel \cite{GiulianiVignale,Ullrich2018}:
\begin{equation}
  \frac{4}{n^2}\! \!\int\limits_0^{h(q)} \frac{q'dq'}{(2\pi)^2} \int\limits_0^{2\pi} \! d\theta \: v_{q'}
  \left(\SC_{\sigma\bar\sigma,\sigma\bar\sigma}^{(0)}(\bfq - \bfq') - n_{\bar\sigma\bar\sigma}\right) = f^{\rm x,S}_{\sigma\bar\sigma,\sigma\bar\sigma}(q)
\end{equation}
The integration limit $h(q)$ can be determined numerically using standard root finding algorithms.
We then use the same integration limit for the non-scalar, full STLS kernel, Eq. (\ref{eq:fxcspinSTLSforce}).
Our modified STLS scheme is schematically illustrated in Fig. \ref{fig:STLSscheme}. In the following, all spin-wave results are
obtained with the so defined full STLS kernel.

\begin{figure}
  \includegraphics[width=\linewidth]{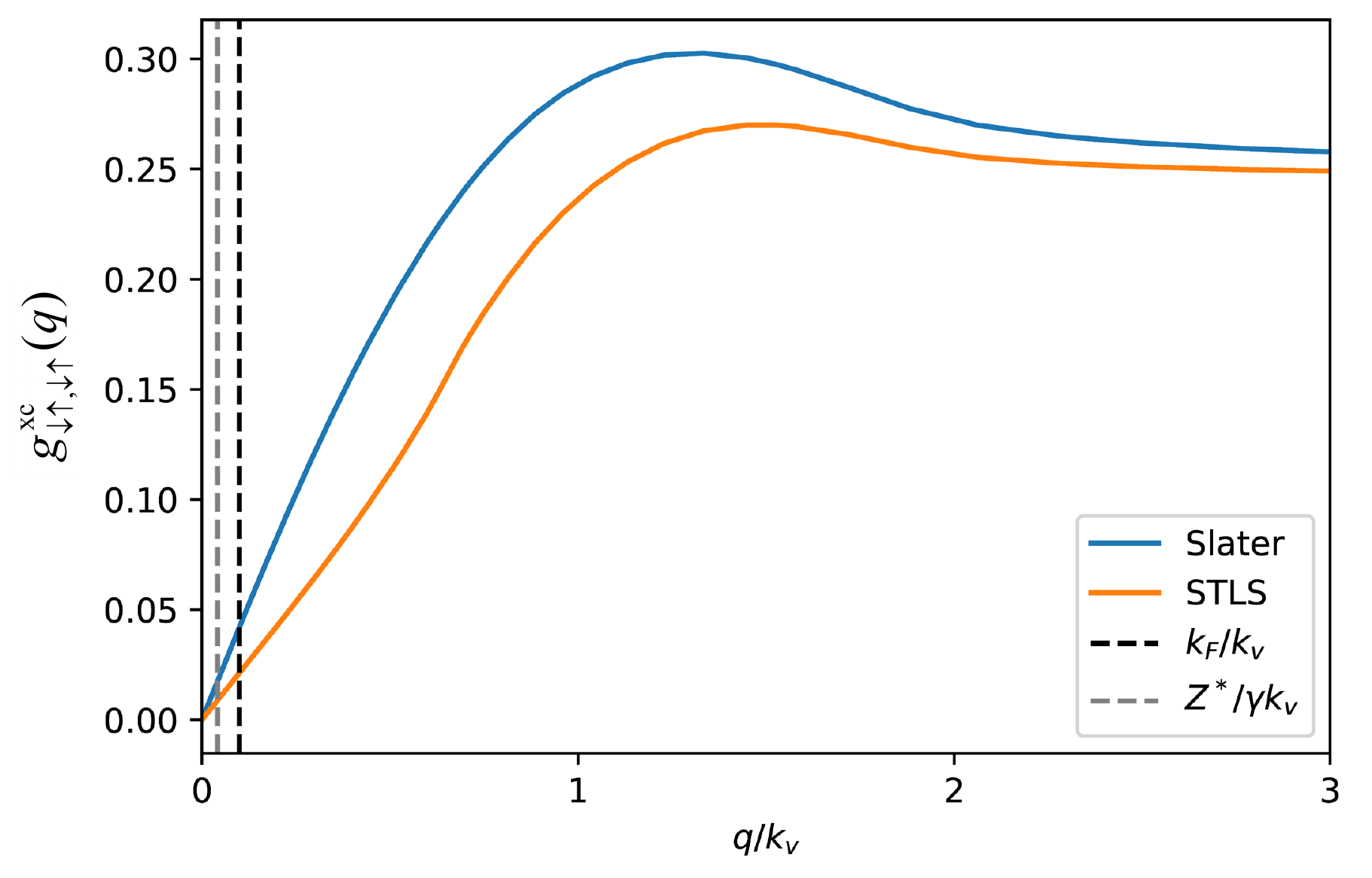}
  \caption{Slater and STLS transverse-spin local field factors $g^{\rm xc}_{\da\ua,\da\ua}(q)$ for $n = 1.89 \times 10^{13}\: \rm cm^{-2}$ and $\zeta = 0.82$.
           The region in which spin-waves can exist is left of the dashed grey line corresponding to the wavevector $Z^*/\gamma = |\ku-\kd|$.}
  \label{fig:LocalFieldExample}
\end{figure}

Figure \ref{fig:LocalFieldExample} shows the Slater and STLS spin-flip local-field factors, defined via
$f^{\rm xc}_{\da\ua,\da\ua}(q) = -v_q  g^{\rm xc}_{\da\ua,\da\ua}(q)$, for $n = 1.89 \times 10^{13}\: \rm cm^{-2}$ and $\zeta = 0.82$.
The xc kernels are dominated by the scale set by the valence electron density $n_v$.
For the spin-wave dispersions, only the region of small $q$ values is relevant, in which the local-field factors have a linear behavior, as indicated
in the figure by the vertical dashed lines. It can be seen that Slater has a larger slope than STLS, which directly affects the spin-wave dispersions, as we
will see below.

\begin{figure}
  \includegraphics[width=0.8\linewidth]{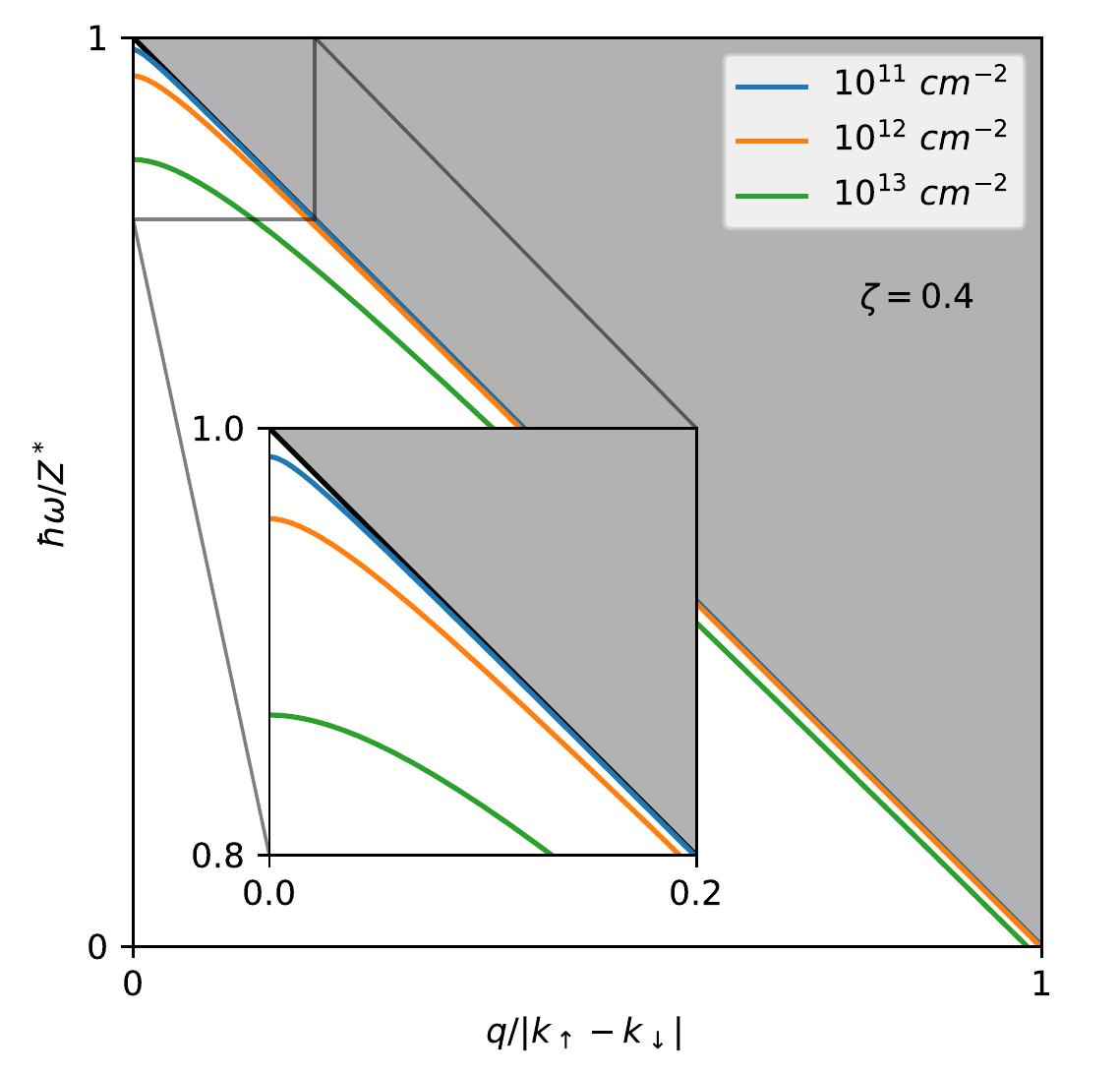}
  \caption{Spin-wave dispersions for various doping densities and polarization $\z = 0.4$.
          The dispersions are scaled by the renormalized Zeeman energy, Eq. (\ref{eq:Zstar}).
          The grey region is the spin-flip continuum.
          The response function has a finite imaginary part in this region and thus the spin-wave damps away.}
  \label{fig:SW-scaled}
\end{figure}

\section{Results and Discussion}\label{sec4}

\subsection{Spin wave characteristics} \label{sec4A}

Figure \ref{fig:SW-scaled} shows spin-wave dispersions, calculated using STLS, for $\zeta=0.4$ and three doping densities:
$n=10^{11}$, $10^{12}$, and $10^{13}\: \rm cm^{-2}$. For smaller densities, the dispersion curves lie closer to the boundary of the
spin-flip continuum; $Z^*-\omega_L$ increases with $n$. The inset to the figure shows a close-up of the spin-wave dispersions for small $q$:
this illustrates how, for smaller $n$, the spin waves are more and more squeezed into a narrow corner below the spin-flip continuum, which
causes the spin-wave stiffness $S_{\rm sw}$ to increase.

\begin{figure*}
  \includegraphics[width=\linewidth]{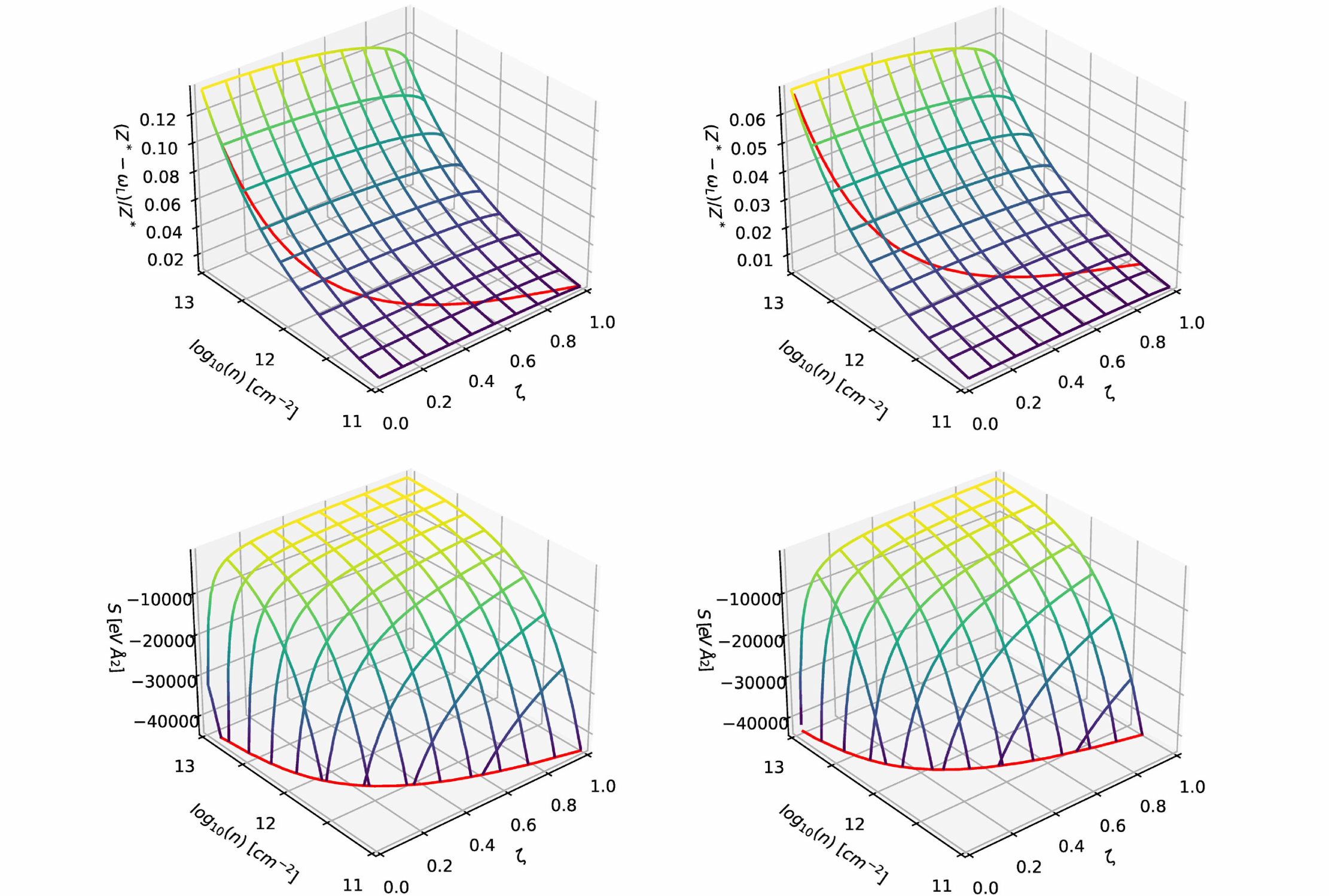}
  \caption{Top left: The pseudo-Larmor frequency, $\w_L$, of spin-waves calculated with the Slater approximation and scaled by the renormalized Zeeman energy, $Z^*$.
           The red line indicates where $Z^*-\w_L = 0.5 \text{meV}$.
           Bottom left: The spin-wave stiffness calculated with the Slater approximation.
           The stiffness diverges as it approaches the origin.
           Right: The same as the left but calculated with the STLS scheme.}
   \label{fig:w0-S}
\end{figure*}

To summarize the characteristic behavior of the spin-flip waves, Fig. \ref{fig:w0-S} shows $1-\omega_L/Z^*$ and $S_{\rm sw}$ as a function of $n$ and $\zeta$,
calculated using Slater (left panels) and STLS (right panels). The quantity $1-\omega_L/Z^*$  represents the $q=0$ offset of the spin wave with respect to $Z^*$,
i.e., the position of the Larmor mode with respect to the spin-flip continuum. Large values of $1-\omega_L/Z^*$ indicate that the Larmor mode is well separated
from the continuum, which increases its lifetime and the chance of it being experimentally observable (see the discussion below).
The red line in the top panels of Fig. \ref{fig:w0-S} indicates that $Z^*-\w_0 = 0.5$ meV, which is comparable to typical linewidths of spin waves found in 2DEGs
\cite{Baboux2012}.

Comparing Slater and STLS, we find that the STLS spin waves tend to lie significantly closer to the continuum than Slater. This is because
exchange is negative and correlation gives a positive correction. We also see this from the slopes in Fig. \ref{fig:SW-scaled}.

The associated spin-wave stiffnesses $S_{\rm sw}$ are shown in the lower panels of Fig. \ref{fig:w0-S}. The stiffnesses in Slater and STLS are
very similar. We find that $S_{\rm sw}$ diverges as $n$ and $\z$ approach zero.
This is because the window in which the spin-wave can exist is shrinking: $0 < \w < Z^*-\gamma q$.
The curvature must get larger in order to fit in this window. At some point this window shrinks to the point of physical irrelevance, which
implies that the spin wave merges with the continuum and ceases to exist as a well-defined collective mode; however, it will still show up as a
resonance feature that can be distinguished from the broad background of the continuum.
We also mention that it is, in principle, possible to observe positive values of $S_{\rm sw}$; however, these would be  for much larger values
of $n$, where the Dirac model is no longer applicable.

\subsection{Prospects for experimental observation}\label{sec4B}

For the experimental observation of spin waves in graphene, doping concentrations
of order $n\sim 10^{11}$ to $10^{13}$ $\rm cm^{-2}$ and significant spin polarizations $\zeta$ are needed. In Appendix \ref{appB} we show that for free-standing
graphene this would require the applications of in-plane magnetic fields of tens to hundreds of Tesla,
which is clearly not practical. Instead, suitable values of $n$ and $\zeta$ should be attainable by depositing graphene on a magnetic substrate.
For instance, experimental and theoretical studies of graphene on Ni(111) have shown that interfacial hybridization of graphene atoms with the top interface
atoms of the magnetic layer causes a spin polarization in the graphene layer \cite{Varykhalov2008,Abtew2013,Peralta2019}. Similarly,
Wang {\em et al.} \cite{Wang2015} demonstrated proximity-induced ferromagnetism in graphene/YIG structures. Wei {\em et al.} \cite{Wei2016}
observed strong interfacial exchange fields (in excess of 14 T) in graphene/EuS structures, with the potential to reach hundreds of Tesla;
device properties may be further improved by encapsulation of the graphene sheet in hexagonal boron nitride \cite{Wang2020}.

Assuming, then, that the necessary conditions (doping and spin polarization) can be achieved in graphene,
the next question is how to create and detect spin waves.
For 2DEG systems in semiconductor quantum wells, spin-flip excitations and spin waves have been studied using
inelastic light scattering (also known as electronic Raman scattering)
\cite{Jusserand1992,Jusserand1993,Perez2007,Perez2009,Baboux2012,Baboux2013,Perez2016,Kung2017}.
For this technique to work, the presence of spin-orbit coupling in the material is essential; clearly, this
rules out pristine graphene, where the spin-orbit coupling is very small \cite{Neto2009}.
Proximity-induced Rashba-type spin-orbit coupling in graphene has been well documented in the literature \cite{Dedkov2008,Wang2016,Avsar2020}.
However, for light-induced spin dynamics, ${\bf L}\cdot {\bf S}$-type spin-orbit coupling in deeper valence bands is needed to enable
spin mixing of interband electron-hole pairs in the 1 eV energy range. Whether these conditions can be achieved by proximity is an open question.
Alternatively, one could excite the magnetization dynamics in the ferromagnetic proximity layer and in this way trigger the
spin dynamics in graphene. However, the resulting hybrid spin modes are expected to be more complex than the pure spin waves considered here, requiring a
theoretical description beyond the model considered in this paper.

An alternative approach could be to use microscopy. Plasmons in graphene have been studied using near-field microscopy \cite{Brar2010,Chen2012,Fei2012}.
Spin-sensitive scanning probes such as spin-polarized scanning tunneling microscopy (SP-STM) \cite{Wiesendanger2009} have been used to
probe spin structure and dynamics of ferromagnets at the atomic scale, including magnon excitations \cite{Balashov2006,Khajetoorians2011,Gao2018}.
There have been STM studies of the electronic and magnetic properties of quantum Hall states in graphene \cite{Song2010,Miller2010,Kim2021},
and it seems conceivable that similar techniques could be used for spin waves.

\section{Conclusion}\label{sec5}

In this paper we have presented a detailed study of spin waves in doped graphene with in-plane spin polarization, using
linear-response TDDFT. From a (TD)DFT perspective, many-body effects in graphene pose an interesting challenge, since
Dirac fermions do not lend themselves to a treatment using approximate density functionals derived from the homogeneous electron gas.
Thus, we placed some emphasis on the development and implementation of orbital-based functionals, and showed that the Slater and STLS approximations
can be successfully used for the charge and spin dynamics in doped graphene.

We calculated spin-wave dispersions and spin stiffnesses for a wide range of doping concentrations and spin polarizations,
and identified regions where the spin waves are well separated from the spin-flip continuum, which means that they should be
sufficiently long-lived to be observable. Creating and detecting spin waves in graphene is without doubt a significant practical challenge,
and we discussed various experimental techniques that appear promising.

Our calculations are based on the ideal model of free-standing graphene with a given Zeeman splitting. In practice, achieving a
spin-polarized Dirac fermion gas most likely involves interaction with a magnetic substrate, which will also introduce spin-orbit coupling.
To account for these effects, our model can be generalized to include Rashba-type spin-orbit coupling; if the Rashba terms are not
too strong, this will preserve the essential features of the spin waves (as is the case in the 2DEG \cite{Baboux2012,Baboux2013,Perez2016,Karimi2017,DAmico2019}).
On the other hand, if the spin waves in graphene are coupled with spin excitations in the magnetic substrate, more complex hybrid modes
may occur. This will be the subject of future studies.

\acknowledgments

This work was supported by DOE Grant No. DE-SC0019109.

\appendix

\section{Derivation of the noninteracting response functions}\label{appA}

Starting from Eq. (\ref{eq:KSResponse}) we get the definition of the noninteracting response function for a Dirac model. We first consider the nonmagnetic case with $Z^*=0$,
where the single-particle energies are $\varepsilon_{\bfk b} = b\gamma|\bfk|$, and the occupation factors become $f_{\bfk b} = \theta(\ef-\varepsilon_{\bfk b})$.
The labels $j$ and $l$ of the single-particle states are replaced with
$j \to (b,\bfk)$ and  $ l \to (b',\bfkp)$, where $b,b'=\pm1$ are the band indices.
The summation over $\bfk$ implies the substitution
\begin{equation}
  \sum_{\bfk} \to \int\limits_0^\infty \frac{k dk}{(2\pi)^2} \int\limits_0^{2\pi} d\phi_k.
\end{equation}
Setting $\bfkp = \bfk + \bfq$, we introduce the orbital overlap function
\begin{equation}
  F^\beta(\bfk,\bfq)=\frac{1}{2}\left(1+\beta \cos (\phi_{k'}-\phi_{k})\right),
\end{equation}
where
\begin{eqnarray}
  \cos (\phi_{k'}-\phi_{k}) &=& \frac{k+q \cos \phi_k}{\sqrt{k^2 + q^2 +2kq \cos\phi_k}} \nonumber\\
  &=& \frac{\kp-q \cos\phi_{\kp}}{\sqrt{\kpsq + q^2 -2\kp q \cos \phi_{\kp}}} .
\end{eqnarray}
The response function then becomes
\begin{equation}\label{eq:DiracResponseDef}
  \chi^{(0)}(\bfq,\omega) = g_s g_v\sum_{b\bp\bfk}
  \frac{(f_{\bfk b}-f_{\bfk' b'}) F^{bb'}_+(\bfk,\bfq)}{\omega+\en{b}{\bfk}-\en{\bp}{\bfkp}+i\eta},
\end{equation}
where $g_s$ and $g_v$ are the spin and valley degeneracies and $\eta$ is a positive infinitesimal required to preserve causality.
Next, we introduce the complex frequency $z$ and separate the response function based on the $b=\pm1$ terms:
\begin{equation}
  \chi^{(0)}(\bfq,z) = \chi_+(\bfq,z) + \chi_-(\bfq,z) \:,
\end{equation}
where
\begin{equation}
  \frac{\chi_+(q,z)}{g_sg_v}=\sum_{\alpha \beta \bfk}\frac{F^\beta(\bfk,\bfq)}{\alpha z+\varepsilon_{\bfk b}-\varepsilon_{\bfk'\beta}}
\end{equation}
and
\begin{eqnarray}
  \frac{\chi_-(q,z)}{g_sg_v} &=&
  \sum_{\alpha \bfk}\frac{F^-(\bfk,\bfq)}{\alpha z+\varepsilon_{\bfk-}-\varepsilon_{\bfk'+}}\nonumber\\
  &=& -\frac{q}{16 \gamma \sqrt{1-\left(\frac{z}{\gamma q}\right)^2}} \: ,
\end{eqnarray}
where $\alpha=\pm1$ comes from separating the $f_{\bfk b}-f_{\bfk' b'}$ terms and transforming the integration limits of the $\bfkp$ integrals and $\beta=bb'=\pm1$ is the (intra)interband transition.
The $\chi_-$ term is a direct continuation from Ref \cite{Wunsch2006}.

Next, we perform the $\beta$ sum in $\chi_+$ to eliminate difficult terms:
\begin{equation}
  \frac{\chi_+(q,z)}{g_sg_v}=
  \sum_{\alpha \bfk} \frac{1}{2\gamma k}\left(\frac{ 1 - \left(\frac{\alpha \zt + 2k}{q}\right)^2 }{ 1- \frac{\alpha \zt}{q}(\frac{\alpha \zt+2k}{q}) + \frac{2k}{q}
  \cos \phi_k} - 1\right),
\end{equation}
where $\zt = z/\gamma$.
The angular integral evaluates to:
\begin{eqnarray}
  \int\limits_0^{2\pi} \frac{d\phi[1-(a+b)^2]}{1-a(a+b)+b \cos\phi} &=&
  \frac{1-(a+b)^2}{(1-a(a+b))} \nonumber\\
  &\times&
  \frac{2\pi}{\sqrt{1-\frac{b^2}{(1-a(a+b))^2 } }}
\end{eqnarray}
and therefore
\begin{eqnarray}
  \frac{\chi_+(q,z)}{g_sg_v} &=&  -\frac{\kf}{2\pi\gamma}
  +\sum_{\alpha }\frac{1 }{4\pi\gamma }\int\limits_0^{\kf}  dk \frac{1-\left(\frac{\alpha \zt+2k}{q}\right)^2}{\left(1-\frac{\alpha \zt}{q}\left(\frac{\alpha \zt+2k}{q}\right)\right)} \nonumber\\
  &\times&
  \frac{1}{\sqrt{1-\frac{\left(\frac{2k}{q}\right)^2}{\left(1-\frac{\alpha \zt}{q}\left(\frac{\alpha \zt+2k}{q}\right)\right)^2}}}\:.
\end{eqnarray}
The radial integral evaluates to:
\begin{eqnarray}
  \!\int \! \frac{dx(1-x^2)}{\left(1-a x\right)\! \sqrt{1-\frac{\left(x-a\right)^2}{\left(1-a x\right)^2}}}
    &=& \frac{x(x^2-1)-\sqrt{1-x^2}\arcsin x}{2(ax-1)\sqrt{\frac{(1-x^2)(1-a^2)}{(1-ax)^2}}} \nonumber\\
     &=& \frac{1}{2}G(a,x).
\end{eqnarray}
This finally gives
\begin{eqnarray}
  \frac{\chi_+(q,z)}{g_sg_v}
  &=&
  -\frac{\kf}{2\pi\gamma}+\sum_{\alpha  }\frac{ q }{16\pi\gamma } \left. G\left(\frac{\alpha \zt}{q},x\right) \right|_{x=\frac{\alpha \zt}{q}}^{x=\frac{\alpha \zt +2\kf}{q}}
  \nonumber\\
   &=&
   -\frac{\kf}{2\pi\gamma}+\sum_{\alpha  }\frac{ q }{16\pi\gamma }  \left(G\left(\frac{\alpha \zt}{q},\frac{\alpha \zt +2\kf}{q}\right) \right.\nonumber\\
   && \left. - G\left(\frac{\alpha \zt}{q},\frac{\alpha \zt }{q}\right)\right) \nonumber\\
   &=& -\frac{\kf}{2\pi\gamma}+\sum_{\alpha  }\frac{ q }{16\pi\gamma }  G\left(\frac{\alpha \zt}{q},\frac{\alpha \zt +2\kf}{q}\right) \label{eq:ResponseDoped}
\end{eqnarray}

The spin-resolved response functions at finite $Z^*$ follow in a similar way, except that the occupation factors are now spin dependent and that the frequency is
shifted by $\essb$. We substitute  $j \to (b, \si, \bfk)$ and  $l\to (b', \si', \bfkp)$, use the single-particle energies
$\varepsilon_{\bfk b\si}=b\gamma|\bfk|+s_\si\frac{Z^*}{2}$ and the occupation factors
$f_{\bfk b\si} = \theta(\ef-\varepsilon_{\bfk b\si})$, and we define
\begin{equation}
  \zst = \omega + \essb + i\eta \:.
\end{equation}
The response function then becomes:
\begin{eqnarray}
  \frac{\chi^{(0)}_{\si\sib,\si\sib}(\bfq,\w)}{g_v} &=&
  \sum_{\bfk b b'}    (f_{\bfk b \si}-f_{\bfkp b'\sib})\nonumber\\
  &\times&
    \frac{F^{bb'}(\bfk,\bfq)}{\w+\varepsilon_{\bfk b\si}-\varepsilon_{\bfkp b'\sib}+i\eta}.
  \end{eqnarray}
To do the $\bfk$-integration we then follow the same procedure as in the non-spin-polarized case above,
taking care to note the different spin occupation factors; in this way, we
arrive at Eq. (\ref{eq:NIResponseSpin}).

\begin{figure}
    \includegraphics[width=\linewidth]{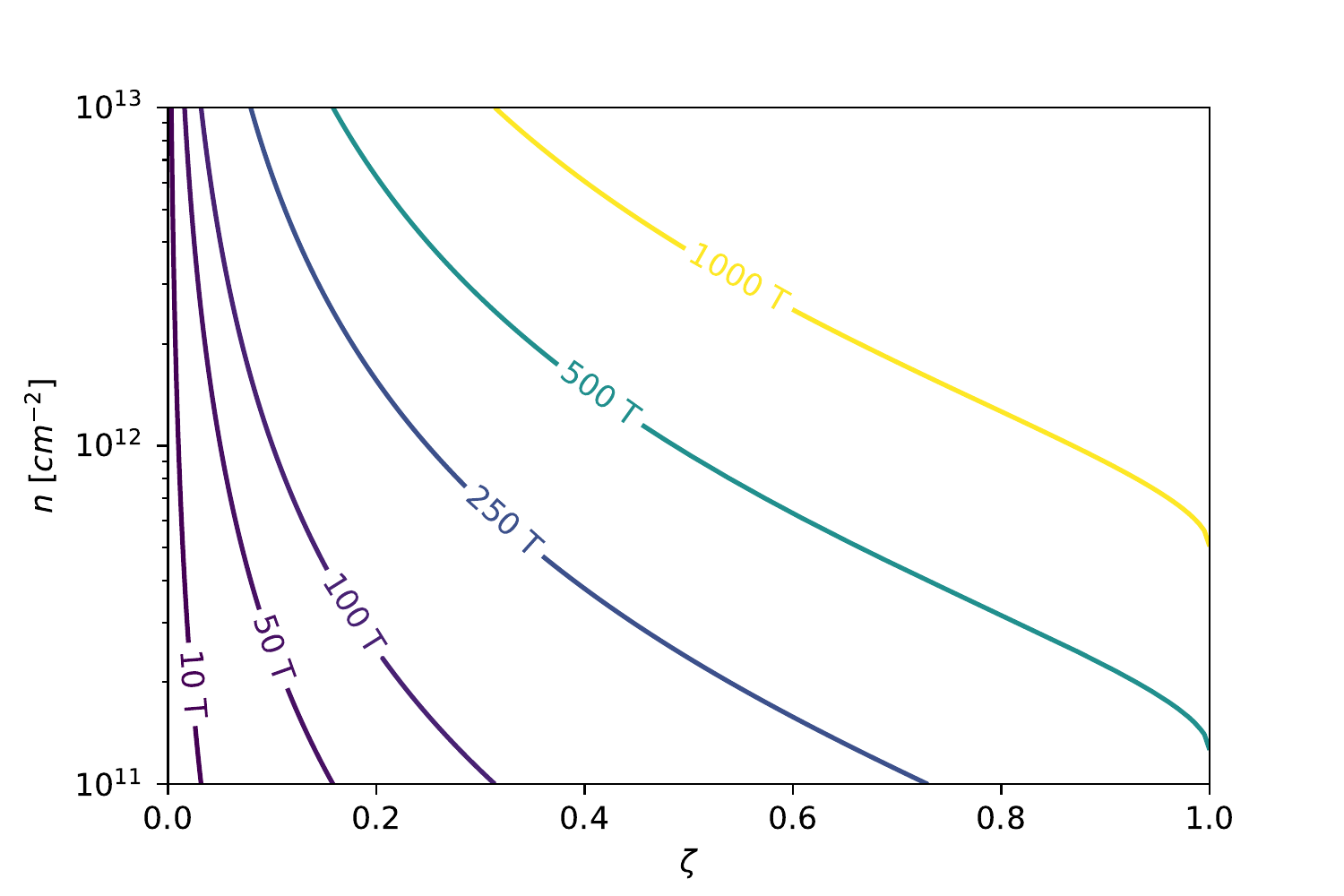}
    \caption{Contour lines of constant magnetic field in the $n-\zeta$ parameter space, illustrating that rather large magnetic fields
    are required to generate significant spin polarization in graphene via the Zeeman effect.}
    \label{fig:MagSplit}
\end{figure}

\section{Magnetic field estimates}\label{appB}

The effective Zeeman energy can be written as
\begin{equation}
    Z^*= g \mu_B (B_{\rm ext} + B_{\rm xc}) =  g \mu_B B_{\rm eff} \:,
    \label{eq:MagSplit}
\end{equation}
where $\mu_B$ is the Bohr magneton and the effective magnetic field $B_{\rm eff}$ is the sum of the externally applied magnetic field $B_{\rm ext}$ and
an additional magnetic field $B_{\rm xc}$ due to exchange-correlation many-body effects \cite{DAmico2019}.
Using the experimental g-factor of graphene, $g=1.952$ \cite{Lyon2017}, we can calculate the $B_{\rm eff}$ that produces
a given value of $Z^*$. Using Eq. (\ref{eq:Zstar}) we can then relate $B_{\rm eff}$ to the spin polarization $\zeta$ and
doping concentration $n$.

This is illustrated in Fig. \ref{fig:MagSplit}, which shows lines of constant $B_{\rm eff}$ in the $n-\zeta$
parameter space. Clearly, a high degree of spin polarization in strongly doped graphene would require very large field strengths.
Notice that $B_{\rm xc}$ is not available in our tight-binding model. Therefore, we cannot obtain the external magnetic field $B_{\rm ext}$
that produces $\zeta$ for a given $n$; however, $B_{\rm eff}$ provides a reasonable estimate for $B_{\rm ext}$ since xc effects can be
expected to be comparatively small.

\section{Some numerical details}\label{appC}

\subsection{Nonuniform $\bfq$-grid}

It is important that our choice of grid spacing for $q$ be sensitive to all of the relevant scales in the model,
determined by the three characteristic wavevectors $k_F$, $k_v$, and $|\ku-\kd|$. It is also important for the $q$-grid to extend all
the way to infinity to account for the integration limits in Eqs. (\ref{eq:fxcspinSTLSscalar}) and (\ref{eq:fxcspinSTLSforce}).
We satisfy these conditions through the repeated use of the following transformation:
\begin{equation}
   t = \left(1+\frac{q}{k}\right)^{-1} \:,
\end{equation}
where $k$ is one of the aforementioned wavevectors, and $q/k \in [0,\infty)$ maps to $t \in (0,1]$.
We then create a uniformly spaced $t$-grid and transform back to a nonuniform $q$-grid.
This results in about half of the $q$ points lying below $k$ and the remaining points having a successively larger spacing.
Finally, we repeat this procedure for each wavevector and merge all of the grids together.
The integration along $q$ is then carried out using integration routines appropriate for nonuniform grids.

\subsection{Frequency integration}

It is numerically convenient to use an alternate definition of the structure factor.
One can use a special construction of the Cauchy integral theorem to show that the integral in Eq. (\ref{structurefactor}) can be transformed into
\begin{equation}
  \mathbb{S}(\bfr,\bfr') = -\frac{1}{\pi}\int_0^\infty \Re \mathbb{\bbchi}(\bfr,\bfr',iu) du \:.
\end{equation}
This expression for the structure factor, involving integration along the imaginary frequency axis, is numerically much better behaved than
Eq. (\ref{structurefactor}). The reason is that $\Im \mathbb{\bbchi}(\bfr,\bfr',\w)$ has minute details along the $\omega$-axis, whereas
$\Re \mathbb{\bbchi}(\bfr,\bfr',iu)$ is quite smooth away from the real frequency axis.
This transformation is the reason why in Sec. \ref{sec3B} we formulate the noninteracting response function with a fully complex frequency as argument.

\bibliography{bibliography}

\end{document}